\documentstyle[aps,prl,preprint]{revtex}
\begin{document}
\draft 
\preprint{IUCAA-12/97}
\title {Gravitational waves from coalescing binaries: 
Estimation of parameters}
\author{R. Balasubramanian  \and  S. V. Dhurandhar}
\address{Inter-University Centre for Astronomy and Astrophysics, \\
Post Bag 4, Ganeshkhind, Pune 411 007, India}
\date{\today}
\maketitle
\begin{abstract}
We investigate the problem of estimation of parameters of a 
gravitational wave signal from a coalescing binary, at  the 
output of a single interferometric detector. 
We present, a computationally viable statistical
model of the distribution, of the maximum likelihood 
estimates (MLE), of  the parameters. This model reproduces the essential
features of the Monte Carlo simulations, thereby explaining 
the large root mean square errors in the estimates, obtained in numerical
experiments. We also suggest a more pertinent
 criterion to assess the performance of 
the MLE, and find that the MLE performs reasonably. 
We have considered a Newtonian signal embedded in Gaussian noise, 
with a power spectrum typical of the LIGO/VIRGO type of detectors.
The model we have used, however, is quite general,
and robust, and will be relevant to many other parameter estimation problems.
\end{abstract}


Coalescing binaries are relatively clean sources of gravitational
waves which  can be modelled with comparatively few parameters.
Numerical experiments using simulated signal
and noise show, that, the actual errors are more 
than a factor of $3$  larger than their 
Cramer-Rao bounds, \cite{BSD95}, at astrophysically relevant
signal-to-noise ratios (SNRs). The comparision with other less
stringent 
lower bounds \cite{NV96} has also not resolved the discrepancy.
In this paper we present a  statistical model which
reproduces the essential features  of Monte Carlo simulations,
even at a `low' SNR of $7.5$. 

The output of the detector will comprise of data trains, each of
duration $T$ seconds. We assume the number of independent uniformly
spaced time samples in each data train to be $N$ which is to be chosen
such that the Nyquist criterion is met.
The set of all
detector outputs constitutes an N-dimensional vector space $\cal V$. 
The statistics of the noise is given by the joint probability of 
the $N$ components of the noise vector {\bf n}, assumed to be
Gaussian,  and is given by,
\begin{equation}
\label{ndis}
p({\bf n}) = \frac{\exp\left[-\frac{1}{2}\sum\limits_{j=-N/2}^{N/2}{\tilde n^j 
\tilde n^{j*}/\tilde {\cal C}_{jj}}\right]
 }{\left[(2\pi)^N \det\left[\tilde 
{\cal C}_{jk}\right]\ \right]^{1/2} }
,
\end{equation}
where the components are expressed in the Fourier domain (indicated by
a tilde) and $\tilde {\cal C}_{jj} = S_n(j/T)$, where $S_n(f)$ is the
power spectrum of the noise. We
assume, $S_n(f) = \left(f/200\right)^{-4} + 2 \left(1 +
\left(f/200\right)^2\right)$, 
consistent with the initial LIGO.

The noise probability distribution
introduces a natural metric on the vector space $\cal V$  via the scalar
product:
\begin{equation}
\label{scal}
\left\langle{\bf x},{\bf y}\right\rangle = 
\sum\limits_{j=-N/2}^{N/2}\tilde x^j \tilde y^{*j}/\tilde {\cal C}_{jj}.
\end{equation}

The set of signal vectors, ${\bf s}(\bbox{\mu}) \equiv
s(t;\bbox{\mu})$, where $\bbox{\mu} 
\equiv \{\mu_0,\mu_1,\ldots,\mu_{p-1}\}$, is a p-dimensional parameter vector,
will describe a p-dimensional manifold $\cal M$ embedded in ${\cal V}$. (See
\cite{BSD95,Ow96} for an introduction to the use of differential
geometry in gravitational wave data analysis). 
Let the output of a detector $\bf x$, contain a signal 
${\bf s}(\bbox{\check{\mu}})$. 
Then  $\bf x = {\bf s}(\bbox{\check{\mu}})  + \bf n$,
where $\bf n$ is a noise vector drawn from the noise distribution. 
The distance, $D(\bbox{\mu})$ between  $\bf x$ and a point 
$ {\bf s}(\bbox{\mu})$ is given by, $D(\bbox{\mu}) =
\left\langle\bf{x - s}(\bbox{\mu}), \bf{x -
s}(\bbox{\mu})\right\rangle^{1/2}$. The MLE of the parameters is that
point $\bbox{\hat{\mu}}$ on the parameter space which minimises
$D(\bbox{\mu})$. This is equivalent to
maximising the correlation 
$c(\bbox{\mu}) = \left\langle {\bf x},{\bf s}(\bbox{\mu})
\right\rangle$ provided we keep $\left\langle {\bf s}(\bbox{\mu}),
{\bf s}(\bbox{\mu})\right\rangle$ constant.

In the limit of high SNR, ${\bf s}(\bbox{\hat{\mu}})$ will lie
in a small
region around ${\bf s}(\bbox{\check{\mu}})$  on the manifold,
effectively the tangent space to the manifold at that point. 
In this case, the difference, ${\bf s}(\bbox{\hat{\mu}}) -
 {\bf s}(\bbox{\check{\mu}})$ can be satisfactorily approximated as a
linear function of $\bf n$. 
Further, if the parameters form a Cartesian system of coordinates, 
then, they too will be linear in $\bf n$ and the distribution of the 
parameters can be described by a multivariate Gaussian \cite{F92}. 
The covariance matrix of this distribution defines a lower bound on the
errors in estimation and is termed as the Cramer-Rao bound. 

If the global minimum of $D(\bbox{\mu})$ is also a local minimum then,
at $\bbox{\mu}=\bbox{\hat{\mu}}$,
$\partial D(\bbox{\mu})/\partial\mu^a = 0$, which implies,
\begin{equation}
\label{max}
\left\langle{\bf s}(\bbox{\check{\mu}}) - {\bf s}(\bbox{\hat{\mu}}) 
 ,\frac{\partial 
{\bf s}}{\partial \mu^a}(\bbox{\hat{\mu}})\right\rangle = 
-\left\langle {\bf n},\frac{\partial 
{\bf s}}{\partial \mu^a}(\bbox{\hat{\mu}})\right\rangle.
\end{equation}
Geometrically speaking,
$\bbox{\hat{\mu}}$ is a local extremum when the tip of the vector $\bf x$
lies on that $N-p$ dimensional hyperplane ${\cal B}_{\bbox{\hat{\mu}}}$, 
which passes through  
${\bf s}(\bbox{\hat{\mu}})$, and is orthogonal to the tangent
plane at ${\bf s}(\bbox{\hat{\mu}})$.
This hyperplane is the
intersection of the $N-1$ dimensional hyperplanes, 
$T^a_{\bbox{\hat{\mu}}}$, each orthogonal
to a coordinate basis vector ${\partial}/{\partial\mu^a}$. 
Let ${\bf l}_a$ be the normalized coordinate basis vectors at
$\bbox{\hat{\mu}}$, 
and let
$r_a$ be the minimal distance from ${\bf s}(\bbox{\check{\mu}})$ to the
hyperplane $T^a_{\bbox{\hat{\mu}}}$, given by $r_a = \left\langle
{\bf s}(\bbox{\hat{\mu}})- {\bf s}(\bbox{\check{\mu}})
,{\bf l}_a\right\rangle$. 
A schematic illustration of the above is given in Fig. \ref{fig0}.

The probability density for the vector $\bf x$ to lie on 
${\cal B}_{\bbox{\hat{\mu}}}$ will depend only on ${\bf l}_a$ and
$r_a$, and can be written down as,
\begin{equation}
\label{rdis}
p(r_a) = \int_{\cal V}\left[\prod_{a=0}^
{p-1}\delta(\left\langle 
({\bf n} - r_a{\bf l}_a),{\bf l}_a\right\rangle)\right] \ p(n)
\ d^{\scriptscriptstyle N}n.                  
\end{equation}
Substituting for $p({\bf n})$ in the equation above and integrating,
we get,  
\begin{equation}
\label{rdis1}
p(r_a) = {{\exp\left[
-\frac{1}{2}\sum\limits_{a,b=0}^{p-1}\left[C_{ab}^\mu\right]^{\bf
-1}r_ar_b\right]}\over{\left[\left(2\pi\right)^p
\mbox{det}\left[C_{ab}^\mu\right]\right]^{1/2}}} ,
\end{equation}
where, $C_{ab}^\mu = \left\langle{\bf l}_a,{\bf
l}_b\right\rangle^{-1}$. 

Since ${\bf l}_a$ are tangent vectors on $\cal M$, the metric on the
tangent basis is nothing but $g_{ab} = 
\left\langle{\bf l}_a,{\bf l}_b\right\rangle$. We can always select a
basis such that $g_{ab}$ are constants over the manifold. If 
the signal manifold is intrinsically flat we can have a coordinate
basis with the same property.
We will assume that the metric coefficients are constant 
in the $\bbox{\mu}$ coordinate system.

Had the map between $\bbox{\hat{\mu}}$ to $\bf r$ been bijective,
it would have been possible to write the distribution for the
estimated parameters, simply as, $p(r_a(\bbox{\hat{\mu}}))
J(\bbox{\hat{\mu}})$, where, $ J(\bbox{\hat{\mu}})$
is the Jacobian of the transformation from ${\hat{\mu}_a}$ to the
variables $r_a$. A given set of values for ${\bf r} \equiv 
\{r_a\}$ could in general 
correspond to a discrete set of  parameter vectors $\bbox{\hat{\mu}}^{(m)}$. 
The problem now is to fix $\{r_a\}$ and compute $P(\bbox{\hat{\mu}}^{(m)}
| {\bf r})$, which is the probability that a particular
$\bbox{\hat{\mu}}^{(m)}$ will be the global maximum for a given $\bf r$.
We will suggest a simple approximation to determine $P(\bbox{\hat{\mu}}^{(m)} 
| {\bf r})$, in the context of our problem. 

The gravitational wave from a CB system can be characterised in the
Newtonian approximation using four parameters, which are, 
(i) the amplitude of the signal, $A$, (ii) the time of
arrival $t_A$, (iii) the initial phase of the the signal (at $t_A$)
$\phi_0$, and (iii) the Newtonian chirp time $\tau_0$ which determines the time
left for actual coalesence after $t_A$. Since it is convenient to have
dimensionless parameters, we define a new set of parameters $\bbox{\mu}
\equiv \{\mu_b\}_{b=0,\ldots,3} \equiv 
\{A,\  2\pi f_At_A,\  \phi_0,\  2\pi f_A\tau_0\}$, where $f_A$ is the
frequency at $t=t_A$. An expression for the
signal in the Fourier domain can be obtained via the stationary phase
approximation (See \cite{SD91} for details)  as $\tilde s(f;\bbox{\mu}) = A \tilde
h(f;\mu_k)$, with,
\begin{equation}
\label{SPA}
\tilde h(f;\mu_k) = {\cal N} f^{-7/6} \exp \left
[i\sum_{k=1}^3\psi_k(f)\mu_k -i\frac{\pi}{4} \right],
\end{equation}
where, ${\cal N}$ is a normalization constant such that 
$\|{\bf h}\| = \left\langle{\bf h},{\bf h}\right\rangle^{1/2} = 1$,
 and $\psi_k(f)$ are
functons of frequency as given below:
\begin{eqnarray}
\label{psi}
\psi_1(f) &=& f/f_A,\;\;\;\;\;\ \   \psi_2(f) = -1,\nonumber\\
\psi_3(f) &=& \frac{f}{f_A} - \frac{8}{5} + \frac{3}{5} \left(\frac{f}{f_A}\right)^{-5/3}. 
\end{eqnarray}
Henceforth, the indices  $i,j,k$ will take  
values in the range $1-3$, and the indices $a,b$ in the range $0-3$. 

For the Newtonian chirp,
\begin{equation}
\label{lr1}
{\bf l}_0 = {\bf h}(\hat{\mu}_k),\;\;\mbox{and}\;\; 
{\bf l}_k = \frac{\partial {\bf
h}}{\partial\mu^k}\left(\hat{\mu}_k\right)\bigg/\left\|
\frac{\partial {\bf h}}{\partial\mu^k}\left(\hat{\mu}_k\right)
\right\|.
\end{equation}
The coefficients $C_{ab}^\mu = \left\langle{\bf l}_a,{\bf
l}_b\right\rangle^{-1}$ are seen to be independent of coordinates
for the chirp waveform. This is a consequence of the intrinsic
flatness of the chirp waveform. 
In order to eliminate the nondiagonal components of ${\bf C}^\mu$, 
we find an orthogonal
matrix $\bf S$ which diagonalises ${\bf C}^\mu$. This transformation
leaves the amplitude parameter unchanged.  
The new parameters $\bbox{\nu}$  are related by
the relation $\bbox{\nu} = {\bf S}\bbox{\mu}$.
In this coordinate system $C^\nu_{ab} = \delta_{ab}$.
The root mean square errors as calculated from the covariance matrix, 
for the parameters $\nu_k (k=1,2,3)$ at an SNR of 7.5 are 
$\{0.020,0.171,4.049\}$ respectively.

In the $\bf\nu$ coordinate system,
\begin{eqnarray}
\label{r1}
r_0(\hat\nu_k,\hat A) &=& \check A \left\langle{\bf h}(\check{\nu}_k),{\bf h}(\hat{\nu}_k)
\right\rangle - \hat A,\\
\label{r2}
r_j(\hat\nu_k) &=&  \check A\left\langle {\bf h}(\check\nu_k),\frac{\partial {\bf
h}}{\partial\nu^j}\left(\hat{\nu}_k\right)\right\rangle\bigg/\left\|
\frac{\partial {\bf h}}{\partial\nu^j}\left(\hat{\nu}_k\right)
\right\|.
\end{eqnarray}
For a fixed set of values, $\{r_j\}$, we will have multiple
solutions, $\{\hat\nu_j\}^{(1)}, \{\hat\nu_j\}^{(2)},\ldots,\{\hat\nu_j\}^{(m)}$,
to eqn. (\ref{r2}). The correlation obtained at these points 
will be $\hat A^{(l)} = \left\langle{\bf
x},{\bf h}(\hat\nu_j)^{(l)}\right\rangle$. (It is to be noted that
for a {\em fixed}
$\{\hat\nu_j\}$, $\hat A$ will be a Gaussian random variable.)
We will set,
$P(\{\hat\nu_j\}^{(l)}|\{r_j\})$ equal to the probability that, $\hat
A^{(l)}$ will be larger than the correlation at all the other multiple solutions.
The identification of the multiple roots is quite a problem, and so we
make the following approximation. For one of the solutions
$\{\hat\nu_j\}^{(l)}$, corresponding to $\{r_j\}$, the probability 
$P(\{\hat\nu_j\}^{(l)}|\{r_j\})$ is almost unity. 
We assume that $\{\hat\nu_j\}^{(l)}$
will be the one which is `closest' to $\{\check\nu_j\}$. If this is
true, then, we only need to compute the probability that the correlation
at an arbitrary point on the parameter space exceeds the correlation
at the point which shares the same value of $\{r_j\}$ but is `closest'
to $\{\check\mu_j\}$.
We
obtain this solution by linearizing eqn. (\ref{r2}), to get,
\begin{equation}
\label{lin}
r_j =\check A\left\|
\frac{\partial {\bf h}}{\partial\nu^j}\left(\hat{\nu}^{(l)}_k\right)
\right\|\left(\hat\nu^{(l)}_j-\check\nu_j\right). 
\end{equation}
So for an arbitrary $\hat\nu_k$ we follow the following procedure,\\
(i)  determine $r_j(\hat\nu_k)$ using
eqn. (\ref{r2}),\\
(ii) determine $\hat\nu^{(l)}_k$ using eqn.
(\ref{lin}),\\
(iii) determine the probablility for $\hat A 
= \left\langle{\bf
x},{\bf h}(\hat\nu_k)\right\rangle$ to be  greater than 
$\hat A^{(l)} = \left\langle{\bf
x},{\bf h}(\hat\nu_k)^{(l)}\right\rangle$, and,\\
(iv)  Set
$P(\{\nu_j\}|\{r_j\})$ equal to the calculated probability  
and attach it as a weight functon to $p(\bbox{\hat\nu})$.

The SNR is essentially the norm of the signal, which is nothing but
$\check{A}$. 
We assume the value of 7.5 for the SNR and
a detection threshold of 6.5. This effectively means that we
neglect all events with $\hat{A} < 6.5$. Since the amplitude parameter
is not of primary interest to us, we shall integrate the distribution
over $\hat A$.
The distribution $p(\hat{\nu}_k)$ is finally given as:
\begin{eqnarray}
\label{ndis1}
p(&&\hat{\nu_k})= 
\frac{1}{\left(2\pi\right)^2} 	
J\left(\hat\nu_k\right)P\left(\hat\nu_k|r_k\right)
\exp\left[-\frac{1}{2}\sum\limits_{i=1}^3 r_i^2(\nu_k)\right]
\;\mbox{\bf\LARGE$\times$}\nonumber\\
&&\int_{6.5}^\infty
\exp\left[-\frac{1}{2}
\left(\check{A}\left\langle h(\bbox{\check{\nu}}),h(\bbox{\hat{\nu}})
	\right\rangle-\hat{A}\right)^2\right] d\hat{A}.
\end{eqnarray}
The integral in the above formula can
be evaluated in terms of the error function.

We now compare the Monte Carlo results with the distribution given in
eqn. (\ref{ndis1}). 
In Fig. \ref{fig1} we plot the marginal 
distribution, $p(\hat\nu_2,\hat\nu_3)$, obtained by integrating
eqn. (\ref{ndis1}) with respect to $\hat\nu_1$.
The surface plot reveals the locations
of the multiple peaks, corresponding to the multiple solutions of
eqn. (\ref{max}). The periodic and regular spacing between the peaks
is a consequence of the intrinsic flatness of the  intrinsic flatness
of the chirp manifold and that the selected coordinates are
cartesian.  We
compare the above with the scatter plot in Fig. \ref{fig2}  
of the estimated values of the parameters on the $\nu_2-\nu_3$ plane, 
where each point corresponds to a realization of noise. The
distribution of the plotted points on the $\nu_2-\nu_3$ plane is
consistent with the marginal distribution. Though we have shown only
the first 2 subsidary peaks, the peaks occur even  beyond the
box shown, but with reduced amplitudes. These peaks are again
consistent with the Monte Carlo results.

 Each peak in Fig. \ref{fig1} 
has a smooth falloff in the $\nu_2$ direction with an approximately
Gaussian profile, but there are sharp cutoffs along the $\nu_3$
direction. This is due to the restriction of the phase parameter to
the interval $[-\pi,\pi]$. The  above effect is also seen in the
scatter plot as the density of points falls of abruptly along the 
$\nu_3$ direction but not along the $\nu_2$ direction. 
Further from the inspection of
Figures \ref{fig1} and \ref{fig2} it is clear that the marginal
distribution $p(\hat\nu_2)$ will show pronounced multimodal behaviour,
whereas  $p(\hat\nu_3)$ will not. Moreover, since the parameters
$\bbox{\mu}$ are linear combinations of $\bbox{\nu}$ the multimodal
nature might not be apparent in the $\bbox{\mu}$ parameters, as can be
observed in \cite{BSD95}. 

We next, compare the one dimensional marginal distribution $p(\hat\nu_1)$
with the histogram obtained via the Monte Carlo method. 
Though the parameter $\nu_1$ has the
least root mean square error of $.02$ 
as predicted by the covariance matrix, its distribution
has the most pronounced non-Gaussian behaviour. In plot (a) and (b) of
Fig. \ref{fig3} we display the distributions $p(\hat\nu_1)$,
obtained from the statistical model and Monte Carlo simulations
respectively. 
Plots (c) and (d) in the same figure zoom in on the 
first two maxima occurring after the central maximum. 
It can be seen that in the case of plot (c) the match is 
not very good even though
the location of the peaks match fairly well. 
The discrepancy here could come from
either the Monte Carlo method or the statistical model. The chief problem
seems to be the use of the linearized equation to determine
$\bbox{\hat\nu}'$. Implementing a more accurate 
inversion procedure close to $\bbox{\check{\nu}}$ would be beneficial. 

Since the distribution of the estimated parameters is multimodal,
using the variances as an indicator of performance of the MLE, 
is not justified. A more  reasonable indicator
would be to compute the probability that the estimated parameters lie
within a certain region around the actual value. 
As a concrete example, we count
the number of times ($N_R$) the errors in the estimated parameters in a numerical
experiment are less than  four times the r.m.s. 
values as computed via the covariance
matrix, out of a total, $N_{tot}$ of points.  
For the Newtonian waveform, at an SNR of $7.5$ 
we find that $90\%$ of the points satisfy
the above criterion. At higher SNRs of $10$ and $15$ the number of
points within the region increases to $95.7\%$ and $99\%$ respectively.
We also applied this criterion for the results of
simulations carried out for the first post-Newtonian waveform 
earlier in \cite{BSD95}, the results of which are given 
in Table \ref{tab1}. 

We see that the MLE works moderately well even at low SNRs.
It is to be remarked that the assessment of an estimator depends upon
how we use the estimates to calculate astrophysically interesting
quantities. In computing the above probabilities we assume equal
importance to all parameters. The above gives only a reasonable but
general measure of the goodness of the MLE. 

We required about 2
days of compution on a 300 MFlops (peak rating) machine to generate
the results of this paper. 
The use of an integration routine specifically
suited to the integrand, and/or the use of lookup tables for computing
the integrand, would further speed up the computation substantially.
Also, for higher values of the SNR the
distribution will be better behaved, and hence, easier to integrate.
The intrinsic flatness of the manifold turns out to be very convenient
for our purpose. This property holds true for the
first post-Newtonian waveform as well, and it will not be
difficult to implement our model for this case.
For higher post-Newtonian corrections and/or for inclusion
of parameters such as spins, it might be computationally expensive to
compute the marginal distributions. However, it is to be noted that
performing Monte Carlo simulations in such cases would also call for a
huge amount of computational effort. 
A further research into the 
above issues is in progress.
 
R.B. would like to thank  CSIR, India for the senior research fellowship.

\begin{table}
\caption{$N_R/N_{tot}$ for the first post-Newtonian waveform.}
\begin{tabular}{c|ccccccc}
SNR&$10$&$12.5$&$15$&$17.5$&$20$&$25$&$30$\\
$N_R/N_{tot}$&$0.797$&$0.872$&$0.922$&$0.956$&$0.976$&$0.980$&$0.988$\\
\end{tabular}
\label{tab1} 
\end{table}
\begin{figure*}
\caption {Schematic illustration of the geometric picture discussed
inthe text. 
}
\label{fig0}
\end{figure*}
\begin{figure*}
\caption { Surface plot of the marginal probability distribution, 
$p(\nu_2,\nu_3)$. The parameter $\nu_2$ is plotted
along the $X$ axis, and $\nu_3$ along the $Y$ axis. The primary peak
has been clipped at a value of $0.1$ at the top. The maximum value
attained was $0.185$. 
}
\label{fig1}
\end{figure*}
\begin{figure*}
\caption { Scatter diagram of the estimated parameters 
on the $\nu_2-\nu_3$ plane. 
}
\label{fig2}
\end{figure*}
\begin{figure}
\caption { Comparision of the Marginal probability distribution,
$p(\nu_1)$ with the histograms obtained from the Monte-Carlo
simulations.   
}
\label{fig3}
\end{figure}
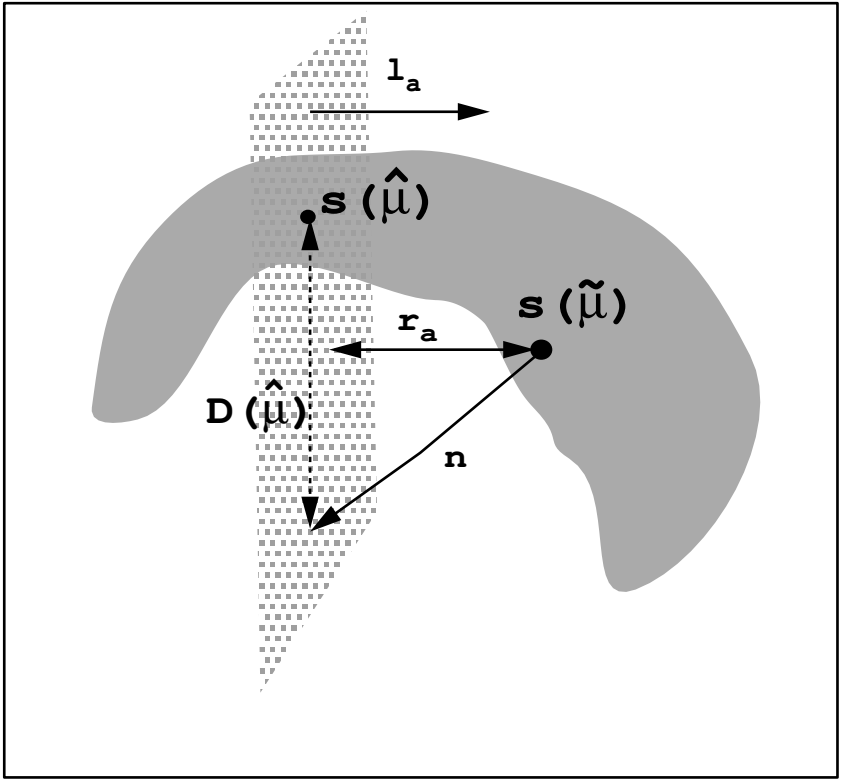
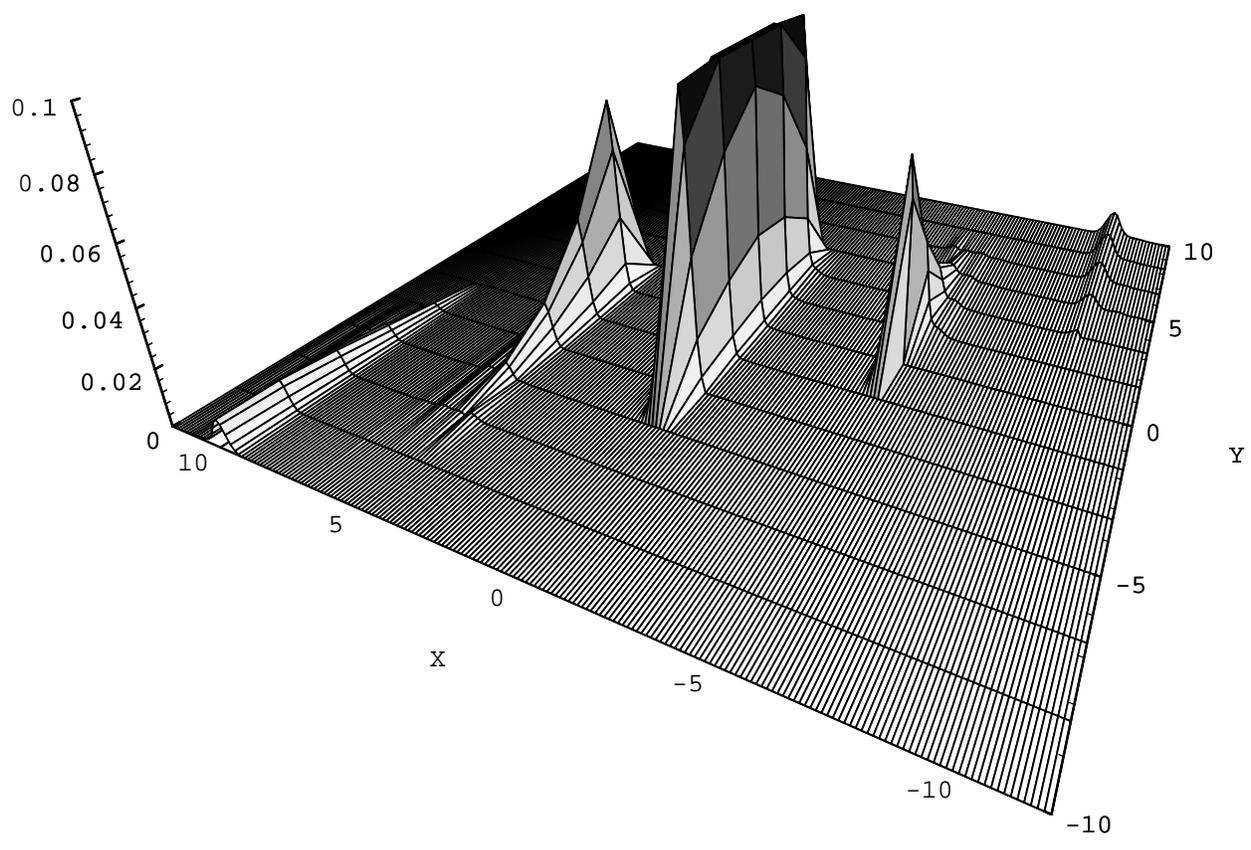
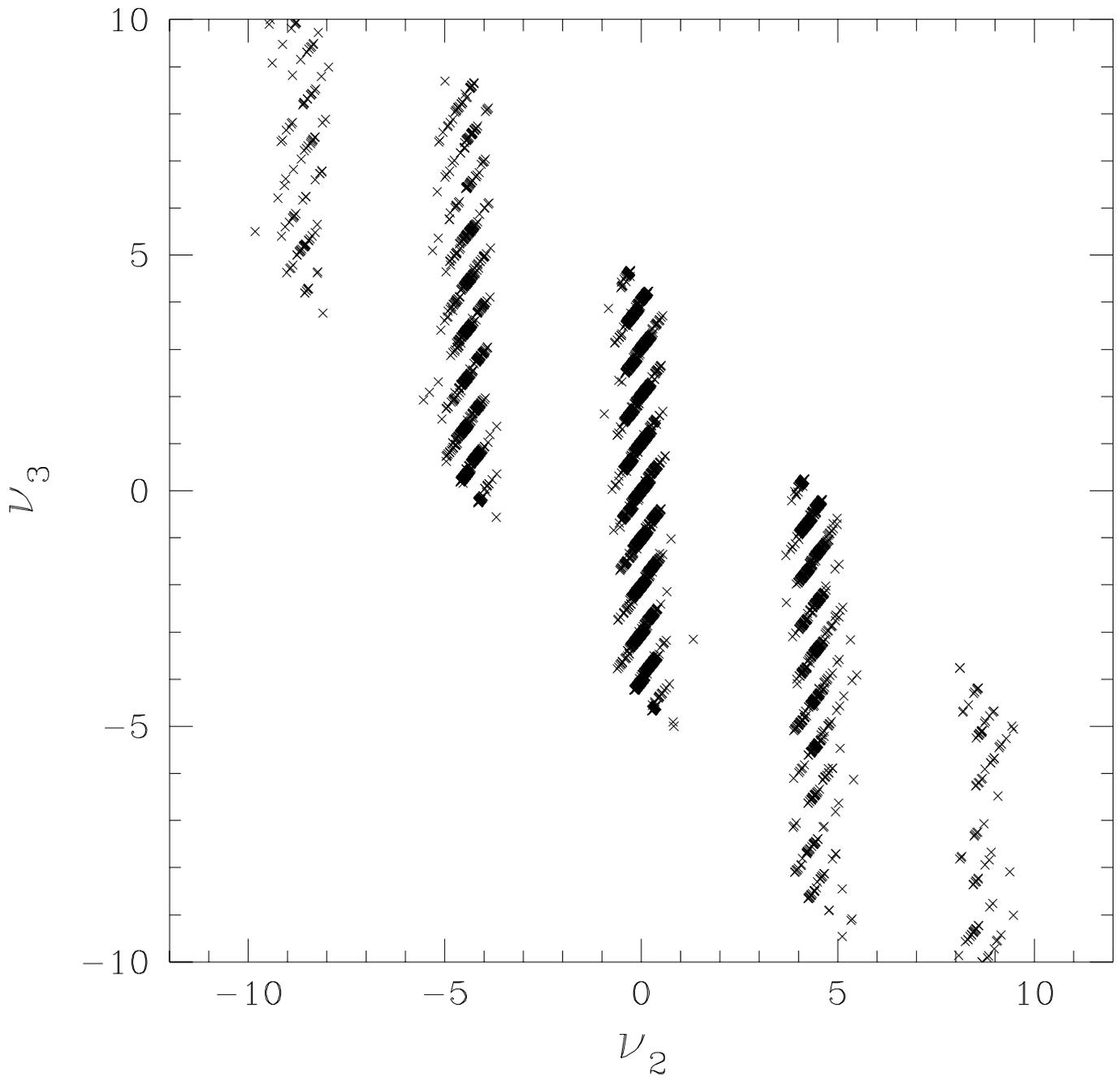
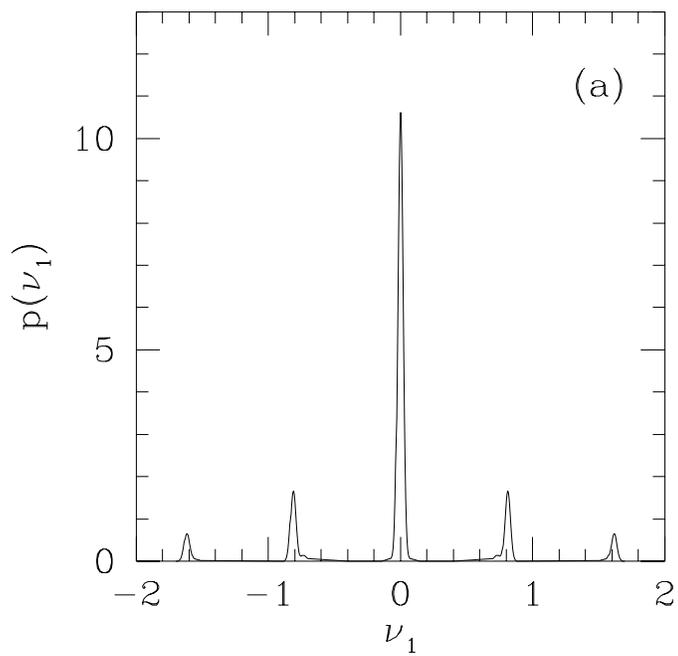
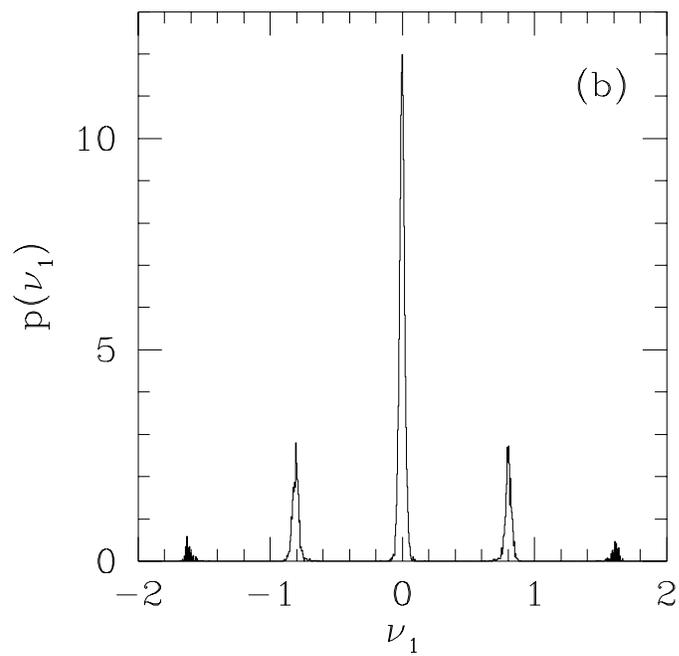
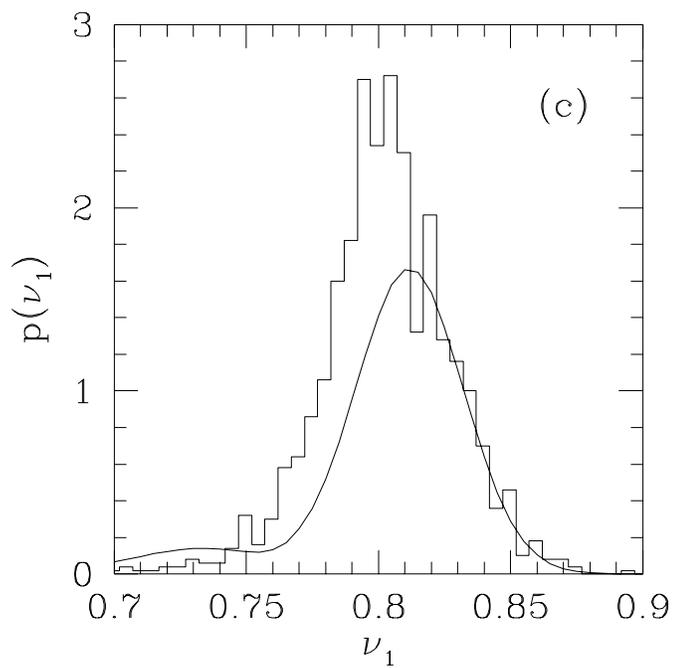
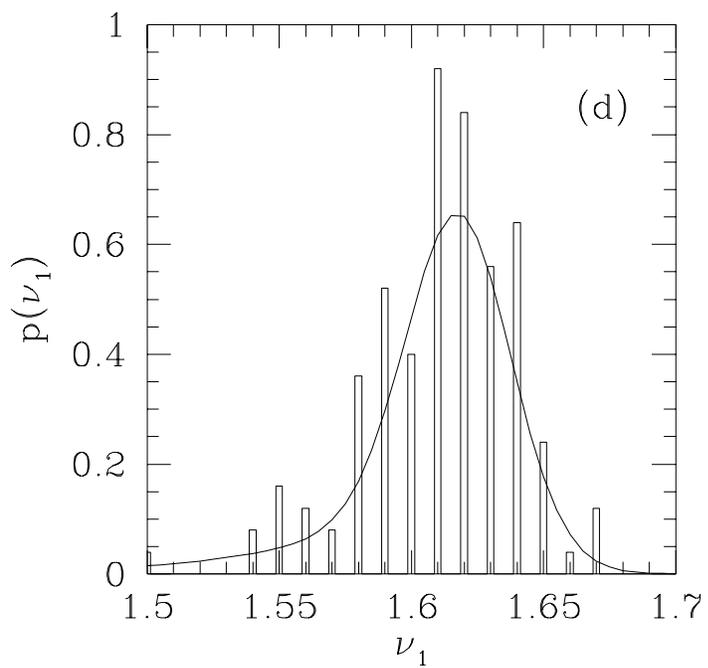
\end{document}